\documentclass[epj]{svjour}
\newif\ifpdf            
\ifx\pdfoutput\undefined \pdffalse \else \pdftrue \fi
\ifpdf 
\usepackage[pdftex]{graphicx}
\pdfcompresslevel9 \DeclareGraphicsExtensions{.jpg,.pdf,.png,.mps}
\else 
\usepackage[dvips]{graphicx}
\DeclareGraphicsExtensions{.eps,.ps} \fi
\usepackage{bm,amssymb}

\newcommand{\beq}{\begin{equation}}
\newcommand{\eeq}{\end{equation}}

\begin{document}

\title{UV-isomerisation in nematic elastomers
as a route to photo-mechanical transducer }
\author{J. Cviklinski, A.R. Tajbakhsh \and E.M. Terentjev}
\institute{Cavendish Laboratory, University of Cambridge,
Madingley Road, Cambridge CB3 0HE, U.K. }

\date{\today}

\abstract{The macroscopic shape of liquid crystalline elastomers
strongly depends on the order parameter of the mesogenic groups.
This order can be manipulated if photoisomerisable groups, e.g.
containing N=N bonds, are introduced into the material. We have
explored the large photo-mechanical response of such an
azobenzene-containing nematic elastomer at different temperatures,
using force and optical birefringence measurements, and focusing
on fundamental aspects of population dynamics and the related
speed and repeatability of the response. The characteristic time
of ``on'' and ``off'' regimes strongly depends on temperature, but
is generally found to be very long. We were able to verify that
the macroscopic relaxation of the elastomer is determined by the
nematic order dynamics and not, for instance, by the polymer
network relaxation.
} 

\maketitle

\section{Introduction}

It has recently been demonstrated~\cite{nishi,hogan_paper} that
properly designed liquid crystalline elastomers (LCE) can exhibit
very strong photo-mechanical effects. Typical LCE are permanently
crosslinked networks of polymer chains incorporating mesogenic
groups either directly into the backbone or as side-groups,
attached end-on or side-on to a flexible backbone. Such materials,
which have been synthetically available for a few years
\cite{fink81}, couple the mobile anisotropic properties of liquid
crystals to the rubber-elastic matrix; many novel and interesting
physical effects arise from this association. For example, the
mechanical shape of a monodomain nematic LCE \cite{kupfer91}
strongly depends on the underlying nematic order, which is a
function of temperature. As the liquid-crystalline assembly
undergoes the nematic-isotropic transition, it loses the
associated anisotropy of polymer backbone chains induced by the
coupling to nematic order. This leads to the macroscopic shape
change of the sample, a uniaxial contraction, which can be
utilised as a thermally-driven artificial muscle \cite{degen}. The
contraction can reach more than $300\%$, with a stress rising up
to $\sim 100\,$kPa, \cite{wermter01,thermal01} and is
theoretically well understood since the late 80s
\cite{gelling,khokhlov}.

\begin{figure}
\centering \resizebox{0.25\textwidth}{!}{\includegraphics{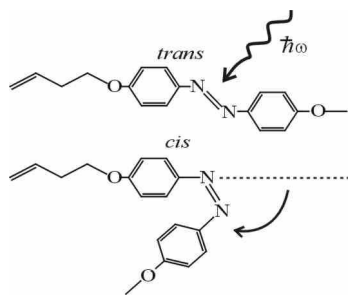}}
\caption{ The effect of azobenzene
\textit{trans}$\rightarrow$\textit{cis} isomerisation on the
molecular shape.  \label{fig1}}
\end{figure}

The nematic order parameter could be manipulated by means other
than temperature, for example by applying an external stress or
electric field, although the latter has been so far shown to have
only a small effect. Another interesting way of affecting the
nematic order is by introducing photo-isomerisable groups in their
chemical structure, so that the mesogenic rod-like molecular
shapes can be contorted by absorption of an appropriate photon.
Several such groups are known in radiation chemistry, e.g.
stilbenes or imines \cite{stilbene,imine}, but the most studied
material in the context of liquid crystals is certainly azobenzene
\cite{azobenzene,eisen78}, Fig.~\ref{fig1}. As their N=N bond is
in the equilibrium \textit{trans} state, the mesogenic moieties
are straight and rigid, and as such contibute to the formation of
the overall nematic order. In the metastable \textit{cis} state
the N=N bond is strongly bent so that the molecular group no
longer has the rod-like shape. Thus, the proportion of
\textit{cis} isomers reduces the nematic order and shifts the
clearing point to lower temperatures, as any other non-mesogenic
impurity would. The resonant frequencies of radiation used to
produce these molecular transformations are approximately the same
in dense polymer systems as in dilute solutions \cite{eisen78}:
$365$~nm for \textit{trans}$\rightarrow$\textit{cis} and $465$~nm
to achieve the back (\textit{cis}$\rightarrow$\textit{trans})
reaction, which also occurs on heating. When azobenzene groups are
incorporated into the polymer chains of a network crosslinked into
a uniaxially aligned nematic elastomer,\footnote{One should also
mention the important work on non-aligned nematic side-chain
azopolymers \cite{keller,ikeda}.} the photo-induced reduction of
nematic order (fully analogous to heating) results in a uniaxial
contraction of the sample, or an increase in force if the sample
is mechanically restricted.

Such photo-mechanical effect in a monodomain nematic elastomer has
been reported for the first time by Finkelmann \textit{et~al.}
\cite{nishi} and studied in some detail by Hogan \textit{et~al.}
\cite{hogan_paper}. However, photo-induced mechanical actuation
has been performed much earlier, using ordinary isotropic polymer
networks \cite{czech,eisen80}, in which the required initial
anisotropy has been induced by load. The magnitude of these
effects was $10^2$-$10^3$ times smaller, with strains of
0.15-0.25\%, and was caused by mechanical contraction of (aligned)
network crosslinks containing azobenzene groups. The large
uniaxial contraction on irradiation of nematic rubbers is a much
more spectacular effect with an entirely different physical
origin.

The main obstacle to turn this effect into a number of appealing
application is its rather slow dynamics \cite{nishi,hogan_paper}.
On irradiation, the response was taking minutes, sometimes hours,
to reach its saturation and it was similarly slow to relax. Could
this be caused by a slow mechanical relaxation in the nematic
elastomer, some time being necessary to change its conformation,
or is it just the photo-isomerisation kinetics that is so slow? At
the first glance, the first option appears quite reasonable
because it is indeed known that nematic elastomers are notoriously
slow in their mechanical relaxation \cite{clarkePRL,hottaJPCM}. In
addition, it is known that although the kinetics of
\textit{cis}$\rightarrow$\textit{trans} isomerisation in dilute
solutions is very fast, in the range of milliseconds
\cite{ikeda2}, the effect in dense polymer melts has always been
reported slow too \cite{keller,ikeda,eisen80}, suggesting the
possible role of polymer dynamics.

To answer this fundamental question, and to try to validate more
quantitatively the models of isomerisation kinetics, we have
studied the photo-mechanical response of monodomain nematic rubber
with azobenzene-containing mesogenic groups in end-on side chains
(thus affecting the nematic ordering, but not having the direct
mechanical effect of network crosslinks \cite{nishi,eisen80}). We
measured the force exerted upon UV-irradiation on the clamps
restricting the sample length. Such a configuration requires no
physical movement of chains in the network, as opposed to the
measurements of changing length of a freely-suspended sample
\cite{nishi,hogan_paper}, when the slow polymer relaxation could
be a greater issue. We, however, found no increase in the speed of
photo-mechanical response. The conclusion that the rate-limiting
process is the photo-isomerisation itself has been further proven
by the simultaneous measurement of nematic order parameter, the
dynamics of which exactly matched the mechanical response.

This paper is organised as follows. The next section gives very
brief description of material preparation, since we use the same
(or very similar) materials to those studied in the pioneering
work \cite{nishi,hogan_paper}, as well as the details of
experimental approach. Section~3 describes the photo-mechanical
response in the ``UV-on'' and ``UV-off'' states, at different
temperatures. Section~4 gives the results of parallel measurement
of the nematic order parameter and relates its magnitude to the
simple linear model of population dynamics of photo-isomerisation.
This data analysis confirms the validity of linear model and
identifies its key parameters. In the Conclusions we summarise the
results and prospects.

\section{Experimental}

\subsection{Preparation of nematic LCE}

\begin{figure}
\centering \resizebox{0.45\textwidth}{!}{\includegraphics{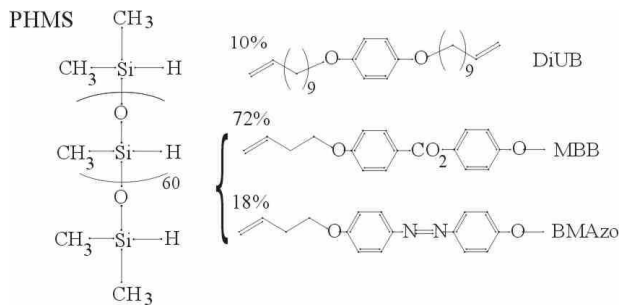}}
\caption{The chemical structures of the compounds used in the
present work, and their mol\% concentrations. \label{fig2}}
\end{figure}

The starting materials and resultant aligned, monodomain nematic
liquid crystal elastomer were prepared in our laboratory. The
procedure for making the side-chain polysiloxanes by reacting the
terminal vinyl group in the mesogenic moiety with the Si-H bond of
the polysiloxane chain, as well as the two-step crosslinking
technique with a uniaxial stress applied after the first stage of
crosslinking to produce and freeze the monodomain nematic
alignment, has been developed over the years by Finkelmann
\textit{et~al}. \cite{kupfer91,kupfer94}. A number of minor
modifications made to the original procedure are described in
\cite{hogan_paper,thermal01}. The azobenzene compounds were all
synthesised according to standard literature techniques
\cite{boden83,vogel,moreuv}. These compounds, and the final
composition of the nematic elastomer, are given in
Fig.~\ref{fig2}. This choice of components and concentrations was
dictated by the results of \cite{hogan_paper}, where the
photo-mechanical response was maximised for this combination. We
consciously avoided placing the azobenzene groups into network
crosslinks, as in \cite{nishi,hogan_paper,eisen80}, in order to
eliminate the direct mechanical coupling to the network: our
purpose here is to study the effect induced purely by the changing
nematic order.

The main mesogenic, rod-like units
MBB\footnote{4'-methoxyphenyl-4-(4''-buteneoxy)benzoate.} are
known to produce a wide-range nematic phase and also reasonably
stable under UV-irradiation (a number of stray photo-chemical
reactions could occur in such complex organic compounds). The
photo-isomerisable rod-like groups were
BMAzo\footnote{[4-(4''-buteneoxy)-4'-methyloxy]azobenzene.}. Both
these end-on side groups have a 4-carbon spacer, which induces a
parallel orientation of the rods to the siloxane backbone
resulting in a prolate chain anisotropy in the nematic phase, with
its principal radii of gyration $R_\| > R_\bot$ \cite{spacer}. The
di-vinyl crosslinking agent is the flexible, non-mesogenic
DiUB\footnote{di-1,4-(11-undeceneoxy)benzene.} which is known to
have only a minor mechanical effect on the nematic network,
preserving soft elasticity and allowing sharp near-critical
evolution on phase transformation. The polymer backbone onto which
all these compounds were grafted was polyhydromethylsiloxane
(PHMS), containing approximately 60 SiH groups per chain, obtained
from ACROS Chemicals. The polymer network was crosslinked via an
hydrosilation reaction, in the presence of a commercial platinum
catalyst COD, obtained from Wacker Chemie. The final composition
of the elastomer is, per Si unit, 72\% MBB, 18\% BMAzo and 10\%
DiUB (the latter figure means that, on average, there are 9
mesogenic side groups between crosslinking points.

Phase sequences were established on a Perkin Elmer Pyris 1
differential scanning calorimeter, which correlated with the
critical temperature obtained by the thermal expansion
measurements and optical microscopy between crossed polars. The
permanently aligned monodomain networks were prepared by the
two-step crosslinking reaction under load \cite{kupfer91}. In the
resulting elastomeric material a broad nematic liquid crystalline
phase was observed, between $T_{\rm ni}=79^{\rm o}$C and the glass
transition at $T_g\approx 0^{\rm o}$C; optical microscopy, X-ray
scattering and mechanical testing confirmed that this was indeed
the nematic phase.

\subsection{Mechanical measurements}

Our mechanical setup is a custom-built device measuring the force
exerted by the sample, and allowing to impose a fixed length and
accurately maintain the temperature of the sample; see
\cite{thermal01,clarkePRL} for more detail. The accuracy of the
force measurements is $\pm$~4.$10^{-5}$~N ($\pm$~0.4~mg).

The rectangular samples ($\sim$10$\times$2.5$\times$0.27~mm) were
mounted with rigid clamps. The thermocouple was then placed behind
the sample, as close as possible. As long as the temperature
variations are slow, this enables us to measure the temperature of
the sample. Of course, concerning the temperature measurements,
the ideal would have been to imbed the thermocouple directly into
the sample during crosslinking, but this would have affected both
mechanical properties and the nematic order. As a compromise, we
placed the thermocouple between the sample and the clamp during
one of the experiments. No difference was found from the case with
a thermocouple close behind the sample.

\begin{figure}[h]
\centering \resizebox{0.5\textwidth}{!}{\includegraphics{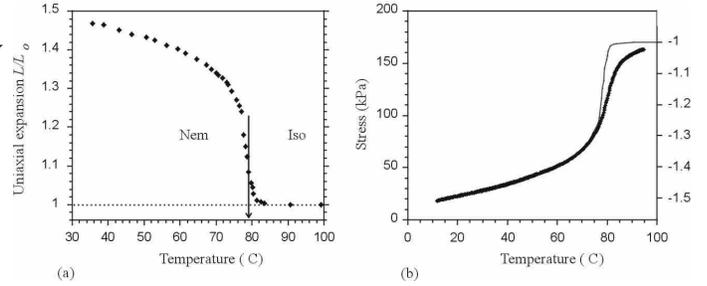}}
 \caption{
Thermo-mechanical properties of the monodomain nematic elastomer.
Plot (a) shows the equilibrium uniaxial expansion/contraction with
temperature between the isotropic and the nematic phase below
$T_{\rm ni}\approx 79{}^{\rm o}$C. Plot (b) shows the stress
arising in the elastomer when it is rigidly clamped at low
temperature and not allowed to contract on heating. The thin solid
line, right $y$-axis, shows the (negative of) strain data from
(a), indicating the direct proportionality of the stress and
strain at low temperatures and strong deviations near $T_{\rm
ni}$. \label{fig4}}
\end{figure}

The UV radiation was produced by a Merck-4.L lamp (power 4W)
providing a narrow band at $365\pm 20$~nm. The window of the
thermally controlled compartment, allowing the passage of the
radiation, was a single standard microscope slide, causing a
$10$\% loss. The intensity of the radiation reaching the sample
was $6$~mW/cm$^2$ (measured by a calibrated photodiode) and kept
constant during the experiments.

The sequence of a typical experiment was as follows:
\begin{itemize}
\item The mechanical history of the sample is first eliminated by annealing, to
start all experiments from the same conditions. The unconstrained
sample was heated to $\sim$120 $^{\rm o}$C and then cooled (slower
than 0.3 $^{\rm o}$C/min) to the ambient temperature.
 \item The temperature was set to a value that will then be kept as
constant as possible; the sample was clamped and equilibrated at
this temperature.
 \item The length of the sample was increased from the natural relaxed
value to a marginally higher value. Thus the force exerted on the
sample increased to a stable value above noise. The sample is
equilibrated at this small extension.\footnote{More precisely, we
wait for the force decay to be slow enough to be ignored. The
complete stress relaxation would probably take weeks, or even be
endless \cite{clarkePRL,hottaJPCM}}
 \item At last, the elastomer is exposed to UV radiation. The force increases
with time and eventually reaches saturation. The force and the
temperature of the sample are constantly logged on the computer.
 \item The source of UV radiations is switched off and the sample is screened
from all light. The force decreases with time and, finally, both
the force and the temperature reach the initial values, set before
the UV irradiation.
 \item After final equilibration, the temperature of the sample
 was given a small variation ($\pm 2-3^{\rm o}$) and the force
 variation logged in order to verify the underlying
 thermo-mechanical coefficient required in the subsequent data
 analysys.
\end{itemize}
The data for the force, transformed into stress by calibrating and
dividing over the (constant) sample cross-section area, and the
temperature were then used in the analysis of the process
kinetics.

\subsection{Birefringence measurements}

The optical birefringence $\Delta n=n_\|-n_\bot$ is one of the
main, and very accurate, measures of the degree of orientational
order in liquid crystals. It was determined using the method based
on measuring the lock-in phase difference $\theta$ between the
split parts of polarised laser beam, one passing through the
birefringent sample, the other through the rotating analyser. The
phase between the two beams is measured by an integer number of
periods, $N$, plus $\theta$, and is related to $\Delta n$ through
eq.~(\ref{phase_eq}), where $\lambda$ is the wavelength of the
laser, $d$ the thickness of the sample.
\begin{equation}
\Delta n= \frac{\lambda}{d} \left( N+ \frac{\theta}{2\pi}\right)
 \label{phase_eq}
\end{equation}
The details of this fast and accurate method are described in
\cite{lim+ho}. In our experiments, we have continuously logged the
values of $\Delta n$ as function of time, before, during and after
UV irradiation. The kinetics of nematic order change, directly
reflecting the population dynamics of \textit{cis} and
\textit{trans} azobenzene isomers, was then compared with that of
the mechanical response.
\begin{figure}[h]
\centering \resizebox{0.48\textwidth}{!}{\includegraphics{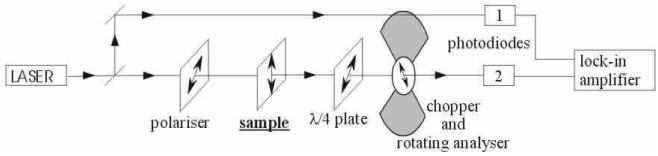}}
\caption{Scheme of differential birefringence measurement,
measuring the phase difference between the two beams, after
\cite{lim+ho}. The free beam (1) is chopped, giving the reference
frequency to lock into; the phase lag of the beam (2) is
proportional to the birefringence $\Delta n$. \label{fig3}}
\end{figure}

\section{Photo-mechanical response}

As long as no radiation is applied, the local nematic order
parameter $Q$ is a function of temperature. If the monodomain
texture is established in an elastomer, then the overall degree of
director alignment has the same magnitude.\footnote{As opposed to,
e.g., polydomain nematic where the local order is $Q$, but
globally the degree of director alignment is zero.} In this case
the elastomer's natural length along the director is determined by
this order parameter. Therefore, this natural length is very
sensitive to temperature variations, see Fig.~\ref{fig4}(a). If,
on the other hand, the sample shape is rigidly constrained, then
instead a force is exerted on the clamps, Fig.~\ref{fig4}(b),
which is in direct proportion to the internal strain due to the
underlying changing natural length \cite{thermal01}. However, one
can notice a difference between the two representations of this
thermo-mechanical effect: the length of the freely suspended
sample, proportional to the nematic order $Q(T)$, changes in an
abrupt near-critical fashion near the $T_{\rm ni}$. In contrast,
the sample clamped at room temperature ends up under increasing
uniaxial stress on heating and, accordingly, the transition to
isotropic phase near the $T_{\rm ni}$ occurs in a diffuse
supercritical fashion.

\begin{figure}
\centering \resizebox{0.4\textwidth}{!}{\includegraphics{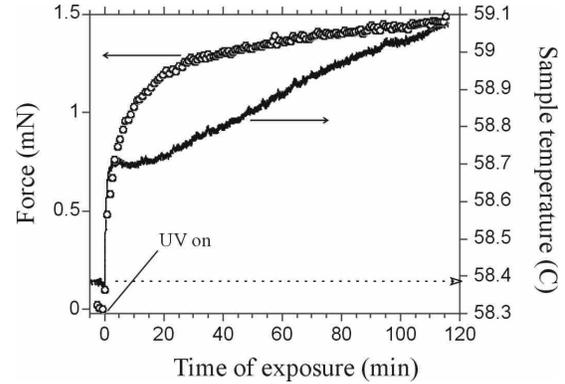}}
\caption{The measured force (open circles, left $y$-axis) and the
measured temperature variation on the sample (scatter line, right
$y$-axis). The initial temperature of $58.4^{\rm o}$C is shown by
the dotted line.  } \label{fig5}
\end{figure}

This strong dependence on temperature has an adverse effect on our
photo-mechanical measurements. Figure~\ref{fig5} shows an example
set of raw data on irradiating a sample clamped in the
dynamometer. As the sample, absorbing UV, was heated by
radiation,\footnote{This was checked by blocking the UV radiation
in the setup remaining otherwise unchanged. In this case neither
temperature, nor force increase has occurred.} a slight
temperature increase (of $\sim 0.5^{\rm o}$C) caused an additional
force increase not related to the photo-isomerisation.

Thus, for more clarity, in the remaining of this paper we present
the results corrected for these unwanted temperature variations.
For the proper temperature adjustment one subtracts from the raw
measured force the value it \underline{would} have without the UV,
at the current (increasing) temperature. To find the latter, we
simply record, independently, the variation of force against
temperature as in Fig.~\ref{fig4}(b).

\begin{figure}
\centering \resizebox{0.32\textwidth}{!}{\includegraphics{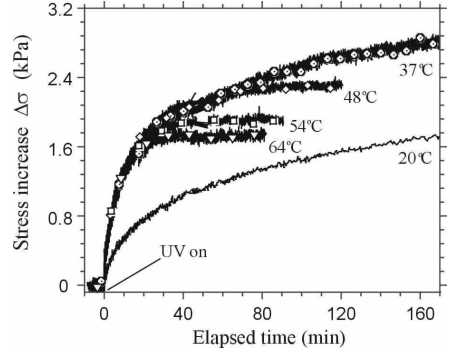}}
\caption{ The stress increase $\Delta \sigma$ caused by the
photo-induced \textit{trans}$\rightarrow$\textit{cis}
isomerisation. The initial temperatures for these experiments are
64.4 , 54.3 , 48.4 , 37.4 and 20.9$^{\rm o}$C, as labelled on the
plot. \label{fig6}}
\end{figure}

\begin{figure}
\centering \resizebox{0.32\textwidth}{!}{\includegraphics{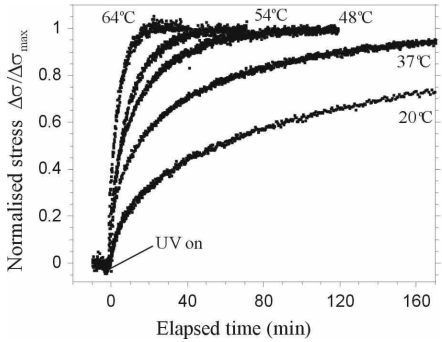}}
\caption{ The normalised stress increase $\Delta \sigma/ \Delta
\sigma_{\rm max}$ caused by the photo-induced
\textit{trans}$\rightarrow$\textit{cis} isomerisation. The initial
temperatures  64.4 , 54.3 , 48.4 , 37.4 and 20.9$^{\rm o}$C are
labelled on the plot. \label{fig7}}
\end{figure}

\begin{figure}
\centering \resizebox{0.32\textwidth}{!}{\includegraphics{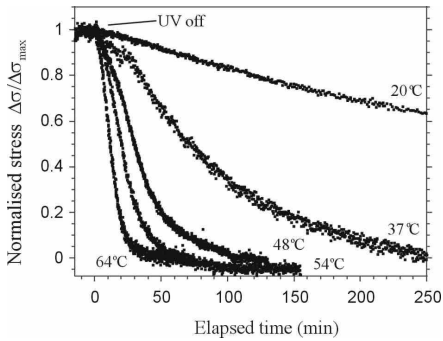}}
\caption{ The normalised stress decrease $\Delta \sigma/ \Delta
\sigma_{\rm max}$ due to the thermal
\textit{cis}$\rightarrow$\textit{trans} relaxation in the dark.
The temperatures are the same as in Figs.~\ref{fig6} and
\ref{fig7}. \label{fig8}}
\end{figure}

The force responses were recorded at various initial temperatures
and adjusted to the fixed value of temperature as described above.
The amplitude of mechanical response depends on temperature, being
more steep in the vicinity of $T_{\rm ni}$, see \cite{hogan_paper}
for more details. Figure~\ref{fig6} shows the set of such stress
increases at different temperatures. The overall magnitude of the
effect is given by the saturation level $\Delta\sigma_{max}$, the
difference between the value the stress finally reached and its
value before the UV to be switched on. The magnitude of
$\Delta\sigma_{max}$ varies over our range of temperatures, but
remains in the range of 1.5-3.5~kPa. To specifically compare the
rates of the response, we plot in Figs.~\ref{fig7} and \ref{fig8}
the normalised results for the increase in stress $\Delta\sigma /
\Delta\sigma_{max}$ on irradiation, and the corresponding thermal
relaxation of stress in the dark.

\begin{figure}
\centering \resizebox{0.4\textwidth}{!}{\includegraphics{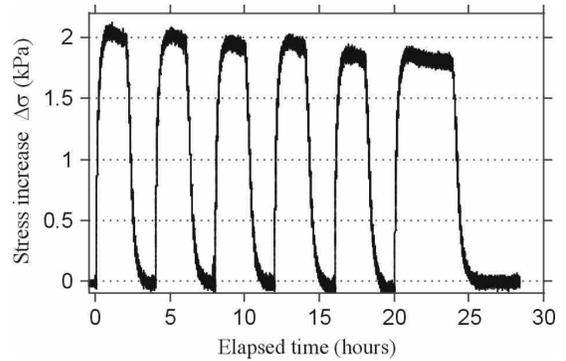}}
\caption{ Stress variations during several light/dark cycles, at a
constant temperature of 56$^{\rm o}$C. \label{fig9}}
\end{figure}

As the stress again reaches its initial value after the UV
radiation is switched off, one could expect the effect to be
repeatable. Figure~\ref{fig9} shows such a test of repeatability,
a crucial characteristics if one to have a practical application
in mind. Overall, the effect is clearly reproducible, however, a
small decay was found when repeating the light/dark cycle many
times. We attribute this weakening to the destruction of few
azobenzene groups (photo-bleaching), after several
\textit{trans}$\rightarrow$\textit{cis}$\rightarrow$\textit{trans}
cycles (one must remember the equilibrium reached upon
UV-irradiation is dynamic, so that the molecules are continuously
cycling from \textit{cis} down to \textit{trans} and back up
again).

\section{Order parameter and population dynamics} \label{section_Q}

The linear population dynamics model presented here is derived
from the one proposed in \cite{hogan_paper}. Despite the good
agreement with experiment, one could argue that it is not the only
possibility. And indeed, at this point, only the final predictions
made in eqs~(\ref{eq_stressup}) and (\ref{eq_stressdown}) were
compared with experimental results. This dynamics could instead be
explained in terms of a mechanical relaxation in the polymer
network, the times measured being in fact those needed for the
polymer chains to respond. So, it was crucial to check
experimentally the main ideas expressed in the simple analysis of
isomerisation dynamics.

\begin{figure} [h]
\centering \resizebox{0.48\textwidth}{!}{\includegraphics{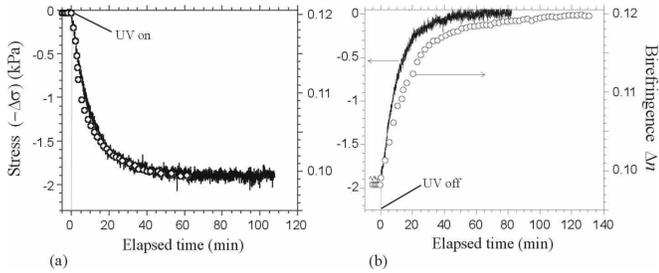}}
\caption{Simultaneous plots of the stress (scatter line, left
$y$-axes) and birefringence (open circles, right $y$-axes),
changing when radiation is applied and then switched off. The
birefringence decrease (increase) reflects the loss (recovery) of
nematic order, which occurs with a rate close to the one of the
mechanical response. The stress variations have been inverted to
superpose the curves and show their similarities. \label{fig10}}
\end{figure}

An answer, albeit qualitative, is given by Fig.~\ref{fig10} in
which both the birefringence and the stress variations ($\Delta n$
and the negative of  $\Delta \sigma$, respectively) are plotted.
The stress variation has been inverted [we actually plot
$\sigma_{initial} -\sigma(t)$] for an illustrating purpose, to
superpose both data sets and show they have \underline{the same}
characteristic dynamics. We do not say the relaxation times are
equal, but they clearly are of the same order of magnitude. This
result is important. First of all it proves the nematic order
(closely reflected by $\Delta n$) actually diminishes upon UV
irradiation. Secondly it shows that this order decreases and then
increases with a speed roughly as the stress variations, and is
therefore the rate-limiting process in this photo-actuation.

As explained in the Introduction, the proportion of \textit{cis}
isomers influences the macroscopic shape of the elastomer through
the decrease in the nematic order parameter $Q$. The
photo-isomerization of small molecular weight azobenzene
compounds, in solution, is known to follow simple first order
kinetics. Only minor modifications were found when the reaction
occurs in a dense polymer matrix \cite{eisen78}. We will therefore
assume, as was done in Ref.~\cite{nishi,hogan_paper}, that the
population dynamics is first order, and more precisely described
by equation
 \begin{equation}
\frac{ \partial}{\partial t} n(t,T) = -\eta \, n -
\frac{1}{\tau_{tc}}n + \frac{1}{\tau_{ct}}(n_0-n) \label{dyn}
 \end{equation}
where $n$ denotes the number of \textit{trans} azobenzene
moieties, $n_0$ the total number of \textit{trans} plus
\textit{cis} isomers (which is constant and given by the initial
composition, Fig.~\ref{fig2}). $\tau_{tc}$ and $\tau_{ct}$ are the
characteristic times of spontaneous
\textit{trans}$\rightarrow$\textit{cis} and
\textit{cis}$\rightarrow$\textit{trans} transformations,
respectively. Unlike the \textit{cis}$\rightarrow$\textit{trans}
transition, the \textit{trans}$\rightarrow$\textit{cis} process
does not occur spontaneously on heating. Accordingly, we shall
neglect $\tau_{tc}$ compared with $\tau_{ct}$. The forced UV
isomerisation occurs with the rate $\eta \equiv 1/\tau_{uv}
\propto$ the radiation intensity.

One can easily obtain the number of \textit{cis} isomers upon
irradiation
 \begin{equation}
n_{cis}(t)=n_0\left(1-\tau_{\rm eff}/\tau_{ct}\right)
\left(1-\exp^{-t/\tau_{\rm eff}} \right) \label{eq_popup}
 \end{equation}
where the light is switched on at $t=0$ and the notation
$$
 \tau_{\rm eff}=\frac{1}{\eta+1/\tau_{ct}} = \frac{\tau_{ct}}{1+\eta \tau_{ct}}$$
is used. Setting $\eta$ to $0$, and assuming in
eqn~(\ref{eq_popup}) that $n_{cis}=n_{cis}(\infty)$ (that is, that
the saturation is reached on irradiation, which is the case
experimentally), the relaxation dynamic is given by
 \begin{equation}
n_{cis}(t)= -n_{cis}(\infty) \exp^{-t/\tau_{ct}}
 \end{equation}
where the light is switched off at $t=0$.

And now, how to take in account the destabilising effect of the
contorted \textit{cis} isomers? As was done in \cite{hogan_paper},
we shall simply consider them as impurities shifting the critical
temperature of nematic-isotropic transition. Thus, as the
mechanical properties of the elastomer are depending on $T-T_{\rm
ni}$, this is equivalent to shifting the actual temperature of the
material. We will assume this shift to be linear,
eqn~(\ref{eq_temp}). This approximation is justified by the rather
small value of the factor (in front of $T-T_{\rm ni}$) this shift
will be found to equal. This factor $\beta$ is positive: the
impurities weaken the order of the nematic liquid crystal:
\begin{equation}
T_{\rm eff}=T_{measured}+\beta \, n_{cis} \label{eq_temp}
\end{equation}

The experimentally measured relation between stress and
temperature, an example of which is given by Fig.~\ref{fig4}(b),
can then be used to obtain the stress variation generated by this
effective temperature shift. The stress-temperature relation, over
the whole range of available temperatures, is of course highly
non-linear. However, again, as the relative effective temperature
shift, or the relative stress increase, are small in our
experiments, one can approximate it as a linear function around
each particular experimental temperature. So, finally, the
following relation stands for the measured stress under
irradiation
\begin{equation}
\sigma(t)-\sigma_{initial}=\Delta\sigma_{max}
\left(1-\exp^{-t/\tau_{\rm eff}} \right) \label{eq_stressup}
\end{equation}
while the relaxation is described by
\begin{equation}
\sigma(t)-\sigma_{initial}=\Delta\sigma_{max}\exp^{-t/\tau_{ct}}
\label{eq_stressdown}
\end{equation}
with $$
 \Delta\sigma_{max}=a \,  \beta n_0
 \left(1-\tau_{\rm eff}/\tau_{ct}\right),
 $$
$a$ being the slope of the discussed linear stress-temperature
relation (which can be easily obtained experimentally) at the
temperature of irradiation. $\sigma_{initial}$ is the stress set
at the beginning of the experiment, before any UV to be applied
(equal in our case to 21.7 kPa for all measurements).

\begin{figure}
\centering \resizebox{0.35\textwidth}{!}{\includegraphics{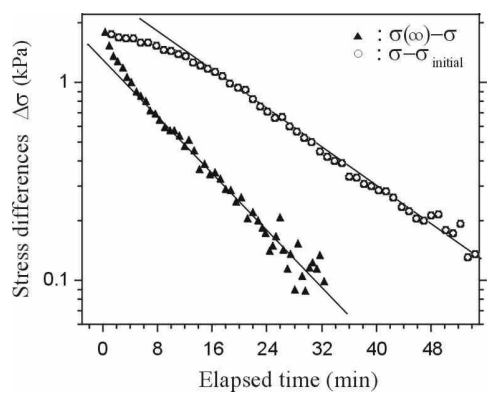}}
\caption{The stress variations with ($\blacktriangle$) and without
($\circ$) UV, in a log-linear scale. The straight lines correspond
to the theoretical predictions of the linear relaxation model, see
text. The isomerisation times thus obtained are plotted in
Fig.~\ref{fig12}. \label{fig11}}
\end{figure}

Does this rather simple model fit our results?
$\Delta\sigma_{max}$ is directly obtained by subtracting the
initial stress $\sigma_{initial}$ from the steady-state saturation
value reached under irradiation. Then, only one free parameter
remains: the characteristic time, $\tau_{\rm eff}$ for the
excitation and $\tau_{ct}$ for the relaxation. An example is given
for both excitation and relaxation in the log-linear plots in
Fig.~\ref{fig11}. If eqs~(\ref{eq_stressup}) and
(\ref{eq_stressdown}) are correct then these curves should be
linear. This appears to be true, except for very short times where
one can notice respectively anomalously fast (at the beginning of
the forced excitation) and slow responses (at the beginning of the
thermal relaxation). This small deviation from the linear model
could be caused by a slightly more complicated isomerisation
kinetics, perhaps dependent on a less trivial shape of the energy
barrier separating the isomer states. But it is not caused by the
actual non-linearity of the stress-temperature relation,
Fig.~\ref{fig4}(b), for explicitly taking into account this
non-linearity does not improve the data fits at all. Whatever the
physical reason, the correction at short times is small and does
not challenge the overall hypothesis of linear dynamics.

\begin{figure}
\centering \resizebox{0.32\textwidth}{!}{\includegraphics{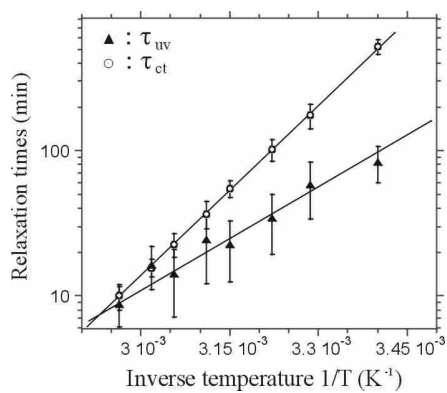}}
\caption{ The characteristic times,  $\tau_{uv}$
($\blacktriangle$) and $\tau_{ct}$ ($\circ$) obtained from the
linear relaxation model fitting; the straight lines are fitting
the data, assuming an Arrhenius behaviour ($\tau\propto \exp
\left(E_{a}/k_{B}T \right)$). \label{fig12}}
\end{figure}

Fitting the time dependence of stress increase on irradiation, or
the decrease on thermal relaxation, provides values of $\tau_{\rm
eff}$ and $\tau_{ct}$ at different temperatures, and thus the
value of the isomerisation rate
 $$\tau_{uv}\equiv  \frac{1}{\eta}=\frac{1}{1/\tau_{\rm eff}-1/\tau_{ct}}.$$
Figure~\ref{fig12} is an Arrhenius plot of these characteristic
times enabling us to estimate the energy barriers for the
isomerisation. We find $E_{a}=$0.5~eV for the light-induced
\textit{trans} to \textit{cis} reaction ($\tau_{uv}$) and
$E_{a}=$0.77~eV for the spontaneous thermal relaxation
($\tau_{ct}$). These times and activation energies are of the same
order of magnitude as those obtained by Eisenbach~\cite{eisen78}
when studying the population dynamics of photo-isomerisable
azobenzene compounds in an ordinary isotropic polymer matrix. In
the framework of this model, we can also calculate the factor
$$
 \beta n_{0}=\frac{\Delta\sigma_{max}}{b
 \left(1-\tau_{\rm eff}/\tau_{ct}\right)},
 $$
as all the terms in the  right hand side are accessible
experimentally. This product, reflecting the ultimate efficiency
of the material as a transducer (assuming $\tau_{\rm
eff}/\tau_{ct}\gg 1$), is equal to 4.0~$\pm$~1~${}^{\rm o}$C,
which is very close to the value found in \cite{hogan_paper} by a
different method.

\section*{Conclusion}

In this work, we have explored the response of a novel
photo-mechanical transducer generating the deformations of forces
many orders of magnitude higher than the effects known before. The
explanations given to the phenomenon, based on the direct effect
of photo-isomerisation on the nematic order parameter and that on
the mechanical state of nematic rubber, have been accurately
validated. We point out that the actuation speed is dominated by
the photo-isomerisation dynamics, which appears to be very slow in
a dense polymer matrix. On the one hand, the isomerisation rate
$\eta$ can be easily increased by simply using more intense
radiation. On the other hand, even leaving aside adverse effects
of photo-degradation, the overall magnitude of the response,
reflected by the effective temperature shift, should be ultimately
limited at $\beta n_{0}$ ($\simeq$ 4~$^{\rm o}$C in our material).

For practical applications, as well as for a better understanding
of the fundamental phenomenon, it would be very interesting to
check these assertions, using various intensities of light. The
maximum temperature shift obtained with our elastomer could seem
small, but can however generate noticeable contractions or forces.
Its intensity should be magnified in the vicinity of the critical
temperature, because of the higher slope of the underlying
thermo-mechanical behaviour there (see Fig.~\ref{fig4}). It would
therefore be useful to synthesise elastomers with transition
temperatures close to the ambient 20$^{\rm o}$C. \\

\noindent This research was made possible due to the Visiting
Fellowship (Stage de Ma\^itrise) of JC, provided by the \'Ecole
Normale Sup\'erieure de Cachan. We thank EPSRC UK for the support
of ART. Invaluable experimental help of S\'ebastien Courty,
Laurent Costier, Atsushi Hotta and Richard Gymer is gratefully
appreciated, as are the discussions with Mark Warner and Yong Mao.


\end{document}